\documentclass[final,3p,times,twocolumn]{elsarticle}
\usepackage[hidelinks]{hyperref}

\usepackage{amsmath,amssymb}
\usepackage[version=4]{mhchem}  
\usepackage{siunitx}
\usepackage{physics}
\usepackage{ulem}
\usepackage{gensymb}
\usepackage[export]{adjustbox}
\usepackage{xparse,mathtools}
\usepackage{comment}
\usepackage[percent]{overpic}

\usepackage{graphicx}
\usepackage{color}
\graphicspath{ {./images/} }
\usepackage{wrapfig}
\usepackage{caption}
\usepackage{natbib}
\setcitestyle{square,numbers}
\usepackage{cleveref}

\usepackage[inline]{enumitem}
\usepackage{subcaption}

\usepackage{booktabs}
\usepackage{multirow}
\usepackage{array}
\usepackage{tabularx}
\usepackage{ragged2e}

\usepackage{listings}
\usepackage{color}
\usepackage{comment}

\usepackage{todonotes}

\begin{document}

\begin{frontmatter}

\title{Current and temperature imbalances in parallel-connected grid storage battery modules}

\author[inst1,inst2]{Joseph P. Ross}

\author[inst2]{Damien Frost}
\author[inst2]{Efstratios Chatzinikolaou}
\author[inst1]{Stephen R. Duncan}
\author[inst1]{David Howey\corref{cor1}}
\cortext[cor1]{Corresponding author} 
\ead{david.howey@eng.ox.ac.uk}

\affiliation[inst1]{
            organization={Department of Engineering Science, University of Oxford},
            city={Oxford},
            postcode={OX1 3PJ},
            country={UK}
            }

\affiliation[inst2]{organization={Brill Power Limited},
            city={Oxford},
            postcode={OX1 2EW}, 
            country={UK}}

\begin{abstract}
A key challenge with large battery systems is heterogeneous currents and temperatures in modules with parallel-connected cells. Although extreme currents and temperatures are detrimental to the performance and lifetime of battery cells, there is not a consensus on the scale of typical imbalances within grid storage modules. Here, we quantify these imbalances through simulations and experiments on an industrially representative grid storage battery module consisting of prismatic lithium iron phosphate cells, elucidating the evolution of current and temperature imbalances and their dependence on individual cell and module parameter variations. Using a sensitivity analysis, we find that varying contact resistances and cell resistances contribute strongly to temperature differences between cells, from which we define safety thresholds on cell-to-cell variability. Finally, we investigate how these thresholds change for different applications, to outline a set of robustness metrics that show how cycling at lower C-rates and narrower SOC ranges can mitigate failures. 

\end{abstract}
\end{frontmatter}

\section{Introduction}
\label{sec:Intro}

Battery energy storage systems (BESS) are rapidly increasing in popularity and cost effectiveness. In the UK, new commercial operations began for 259 MW of storage in Q3 of 2024 alone. Furthermore, with the closure of the last UK coal plant in 2024, long-term revenues are continuing to increase at pace \cite{Modo-OctoberRoundup}. A typical BESS architecture consists of groups of cells connected in series or parallel into a module, and these modules are connected in series to increase the overall pack voltage \cite{RENIERS2023120774}. Although a parallel connection is not always necessary, high-power or long-duration systems require parallel connections to remain in accordance with the International Electrotechnical Commission standard, which recommends that the maximum nominal voltage of the battery stack should remain less than  \SI{1.5}{kV} \cite{IEC63056}. One challenge with parallel-connected systems is that the BMS typically does not sense individual cell currents and temperatures. Due to cost and space constraints, only total module current and just a few temperatures are measured. In this arrangement single-cell failures, as well as current and temperature imbalances, may not be detected by the battery management system (BMS). As a result, BMS measurement and diagnostic tools in multicell systems must overcome the aforementioned sensing challenges \cite{Tanim2021}. It is therefore important to understand the intricate dynamics of parallel systems so that packs can be designed to mitigate these failures and BMS diagnostic tools can ensure safe and long lasting operation without requiring additional sensors. 

\begin{figure*}%
    \includegraphics[width=0.9\textwidth, center]{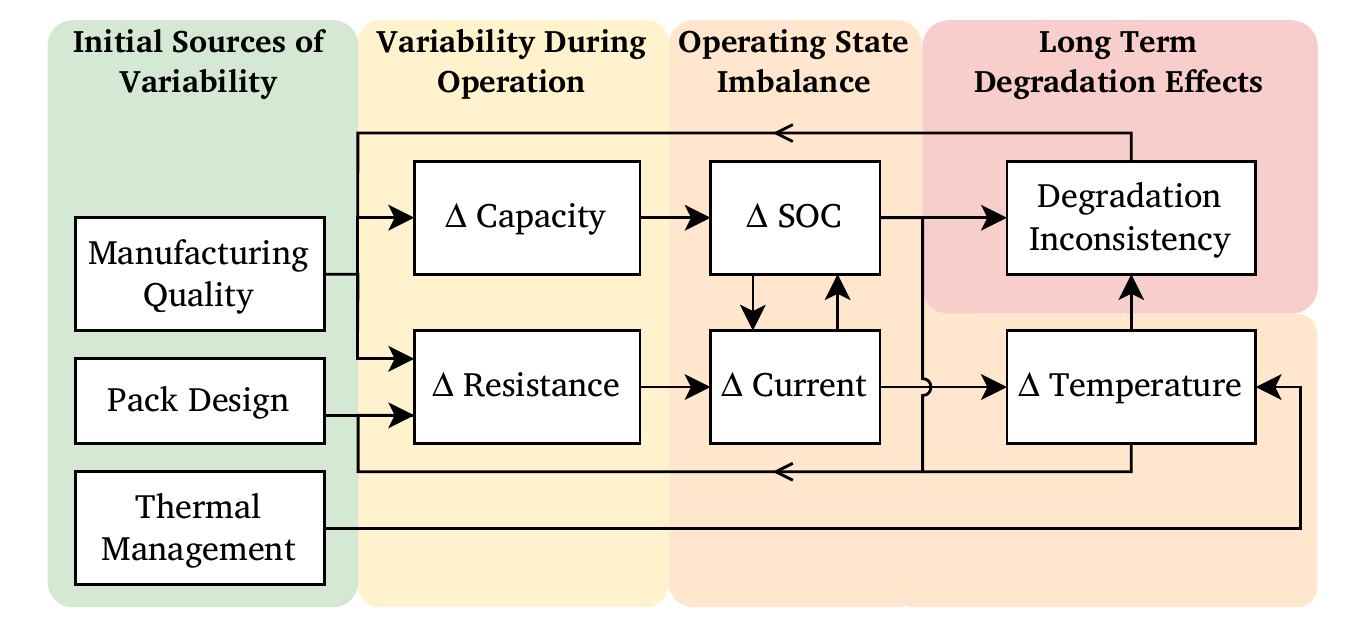}
    \caption{Development and feedback of BESS parameter variation from manufacturing to end of life.}
    \label{paramVariationFlow}
\end{figure*}

In the ideal scenario, two identical cells connected in parallel would be loaded with the exact same current. However, no two cells are perfectly identical; there are small differences in cell impedance and voltage-SOC (state of charge) dynamics that result from manufacturing variability \cite{RUMPF2017224}. This causes parallel-connected cells to be loaded with different currents, and therefore to operate at different SOCs and temperatures \cite{WU2013544, WANG2022104565}. The different operating conditions have several effects: cells degrade at different rates, reducing the useful life of the module; localised degradation and local extreme temperatures pose a safety risk \cite{8626763}; and SOC imbalances mean the full amount of energy cannot be extracted out of each cycle \cite{Zhang2018_CurrentDist}. 

Initial cell-to-cell variability can increase over a module’s lifetime as differences in operating conditions drive uneven degradation. Resistance differences can arise from mismatched degradation rates \cite{BAUMANN2018295, RENIERS2023120774, Reniers_2019}, or cell short-circuit and open-circuit failures \cite{KONG2018358, BIRKL2017373, 9913923, Tanim2021}, resulting from stress such as binder contact loss, separator pore clogging, lithium plating, or current collector dissolution. Module connection resistances also play a role, as cells further from the output terminals have a higher interconnection resistance \cite{WU2013544, hallArray}, and contact resistance failures may occur due to corrosion or mechanical stress \cite{ZWICKER2020100017, HENDRICKS2015113, OFFER2012383}. Differences in cell capacity also result from degradation inconsistencies. Individual cells may lose capacity faster than the wider group due to failures such as electrode mechanical breakdown, passivation layer growth, lithium plating, electrode overhang issues, or accelerated capacity fade (commonly known as the `knee point') from a combination of mechanisms \cite{edge2021lithium, BIRKL2017373, o2022lithium, HARRIS2017589}. The wider impacts of cell-to-cell variability, imbalanced operating conditions, and differing degradation rates has drawn attention in the literature \cite{BAUMANN2018295,YANG2016733, BRUEN201691}, and can be distilled into Fig.\ \ref{paramVariationFlow}. 

The dynamics of these imbalances can create a situation where there is a positive feedback loop between current and temperature imbalances, allowing temperature gradients to persist \cite{FILL2022104325}. From an initially balanced SOC, the cells in a module will be loaded such that the cell with the lowest resistance will have the highest current. This cell will heat up the most, which will reduce its internal resistance, which will increase the current and so on, until a steady state is reached. Under these circumstances, a persistent temperature difference could lead to reduced lifetime and performance \cite{NaylorMarlow2024}.

Current and temperature dynamics in parallel cells have been investigated comprehensively to understand why they occur and their impact on performance and lifetime. However, there exists a research gap connecting these topics; there is only a small body of research that quantifies typical imbalances in an industrial or utility-scale BESS. Researchers have performed experiments and simulations on parallel cells with large applied thermal gradients of up to \SI{25}{\celsius} \cite{NaylorMarlow2024, LIU2019489} and examined the resulting degradation, finding that there is a significant impact on lifetime. Others have measured the performance of parallel-connected cells under more realistic and benign thermal boundary conditions and found that typically smaller temperature differences (less than \SI{2.2}{\celsius}) across a module arise during cycling \cite{PIOMBO2024110783}. While these studies achieve their goals of understanding the origins and impacts of thermal gradients, they have not addressed the general impact of large temperature differences on the safety and lifetime of the system. Here, we study a system particularly susceptible to parallel imbalances in a module constructed from large-format prismatic LFP cells. The results are intended to help system manufacturers understand and improve on the typical limitations of a BESS with parallel connections. 

Prismatic LFP cells, the focus of this investigation, are currently among the cheapest (per kWh) batteries in the stationary storage market. These cells have wound or stacked electrode layers within a rigid aluminium alloy case. This format enables a high capacity ($\geq$ \SI{100}{Ah}) and low impedance ($\leq \SI{500}{\micro\ohm}$) per cell, well suited to high-energy applications such as grid storage. With a recent increase in manufacturing output and a subsequent decrease in production cost, prismatic LFP cells are projected to dominate the stationary storage industry in the near term \cite{bloombergBatteries, BAJOLLE2022102850, ORANGI2024109800}. Although they have an established market share, they are not without reliability concerns \cite{Roy31122024}. The low impedance and high capacity coupled with a flat open circuit voltage (OCV) vs.\ SOC curve makes them particularly susceptible to imbalanced currents and temperature fluctuations when connected in parallel. This poses significant system-level challenges for BESS manufacturers. 

Imbalances between parallel-connected large LFP cells occur for several reasons. Prior studies report that contact resistances can significantly impact module available capacity and performance \cite{PIOMBO2024110783}. A critical parameter is the ratio between contact resistance and cell resistance \cite{HOSSEINZADEH2019194}; this should be less than 2\% to avoid issues \cite{WU2013544}. Cell interconnection resistances should ideally be around \SI{10}{\micro\ohm} \cite{RENIERS2023120774, 10644553}, although welded joints are typically higher \cite{ZWICKER2020100017}. In addition, the inherently flat OCV-SOC characteristic of LFP limits voltage-driven self-balancing between parallel cells, such that small resistance differences are not readily compensated \cite{weng2024current}. This means that differences caused by factors such as contact resistances cannot be corrected easily. Despite this, current and temperature differences in these cells have not been studied extensively. 

A key reason that large parallel-connected modules and large format cells have not drawn significant attention in the literature is because conventional battery cyclers are low-current devices, making experiments time consuming or impossible. Here, we measured the system dynamics with cycling experiments performed on a module of four large cells in parallel. Additionally, we simulated the system with an electrothermal dynamic model, connecting individual cell models in parallel to resolve the branch currents by imposing Kirchhoff constraints. This approach allows us to understand how individual currents depend on other states in the system. The feedback loop between temperature and resistance is also captured by considering resistance to have an Arrenhius dependence on temperature \cite{TROXLER20141018, 6861800}. We fitted the model parameters to cycling data to ensure the model represented an industrial system and gave accurate predictions. A sensitivity analysis on thermal imbalances was performed, and from this we have ascertained safety limits linked to cell-to-cell variability that can be used as BESS module design constraints. 

\section{Model Formulation}
\label{sec:modelFormulation}

\begin{figure*}%
    \includegraphics[width=0.75\textwidth, center]{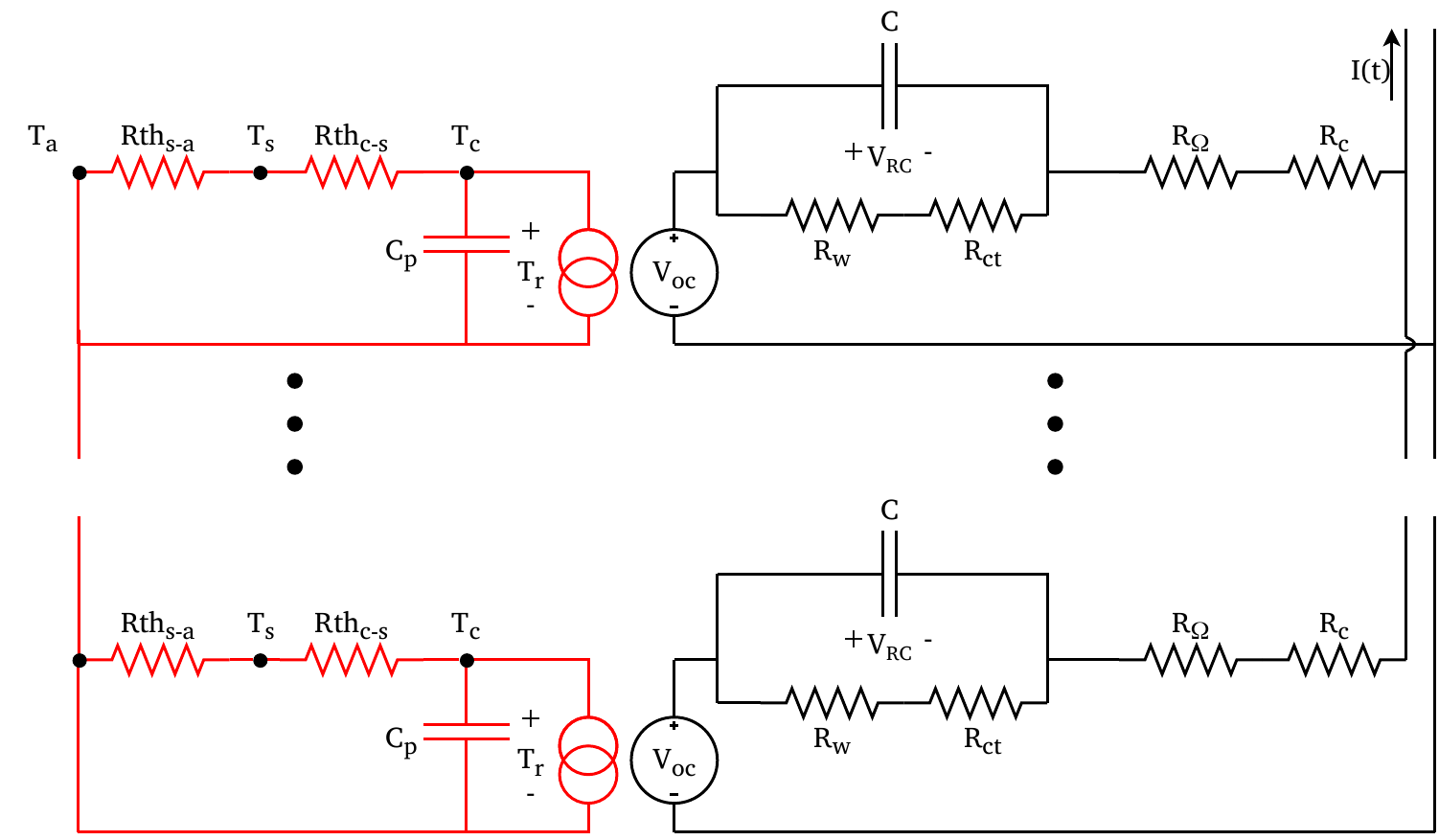}
    \caption{Model electrical-thermal diagram}
    \label{modelDiagram}
\end{figure*}

An empirical nonlinear state-space model of a parallel-connected module was formulated to investigate the interplay between currents and temperatures in parallel cells. In this approach, each cell is represented with a temperature-dependent equivalent circuit model (ECM) to describe the electrical dynamics, and a lumped thermal circuit to describe the temperature dynamics. The cell models are coupled in parallel, with the input to the state-space model being the total module current and the output the surface temperature of each cell. A diagram of the model is shown in Fig.\ \ref{modelDiagram}. 

The ECM for each cell consists of an OCV source, a resistor-capacitor (RC) pair, and two series resistors. (To simplify the derivation of the parallel behaviour, we shall initially ignore the RC-pair, as discussed below.) One series resistor represents the ohmic losses in the cell (\(R_\ohm\)), and one represents the contact resistance (\(R_\text{c}\)). The RC pair also includes two separate resistances: \(R_{\text{ct}}\), that depends on temperature, and \(R_\text{w}\), independent of temperature. 

In models of parallel cells, the dynamics can be described with a system of differential algebraic equations (DAEs), comprising ordinary differential equations (ODEs) that describe the dynamic relationships between inputs and outputs, and one or more algebraic equations that impose Kirchhoff's laws. The DAE model may be turned into an ODE system by resolving Kirchhoff constraints and expressing branch currents in terms of the system states and parameters \cite{9662282}, and this approach has been used to study current imbalances in vehicle battery packs \cite{BRUEN201691}. The benefit of this approach is that the individual branch currents can be resolved using the total current and a known parameter set, which is important in this work because all the model states depend on the individual currents.

We now show how to solve for the branch currents, starting with a DAE representation of a simplified OCV-R model of cells in parallel: 
\begin{equation}
    \begin{bmatrix}
        1 & 0 \\
        0 & 0
    \end{bmatrix}
    \begin{bmatrix}
    \dot{z} \\
    \dot{i}
    \end{bmatrix}
    =
    \begin{bmatrix}
        0 \\
        \Delta f(z)
    \end{bmatrix}
    +
    \begin{bmatrix}
    0 & A_{12} \\
    0 & A_{22}
    \end{bmatrix}
    \begin{bmatrix}
        z \\
        i
    \end{bmatrix}
    +
    \begin{bmatrix}
        0 \\
        B_2
    \end{bmatrix}
    I.
    \label{initialModel}
\end{equation}
For $N$ cells connected in parallel, \(z\) is an $N$-element vector of the SOCs of all cells, \(i\) is an $N$-element vector of the branch currents, and \(I\) is the total current. In general, discharging currents are assumed to be positive. Rearranging the second row in (\ref{initialModel}) to solve for $i$: 
\begin{equation}
    i = -A_{22}^{-1}(\Delta f(z) + B_2I).
    \label{containsCurrents}
\end{equation}
Here, $\Delta f(z)$ is a vector of voltage difference terms, as shown in (\ref{eqfZ_A12}), where $f(z_x) = \text{OCV}_x$, i.e., the OCV for the $x$th cell having SOC of $z_x$. Additionally, continuous-time SOC is defined as the integral of the current, and so $A_{12}$ is a diagonal matrix of inverse cell capacities: 
\begin{align}
&\Delta f(z) = 
    \begin{bmatrix}
        f(z_1) - f(z_2) \\
        f(z_1) - f(z_3) \\
        \vdots \\
        f(z_1) - f(z_N) \\
        0 
    \end{bmatrix},
    & A_{12} = 
    \begin{bmatrix}
        \frac{1}{Q_1} & \dots & 0 \\
         & \ddots &  \\
        0 & \dots & \frac{1}{Q_N}
    \end{bmatrix}.
    \label{eqfZ_A12}
\end{align}

The remaining matrices enforce the constraints so that all branch currents sum to the total current and the terminal voltage is equal across the cells: 
\begin{align}
    &A_{22} = 
    \begin{bmatrix}
        R_1 & -R_2 & 0 & \dots & 0 \\
        R_1 & 0 & -R_3 & \dots & 0 \\
        \vdots & & & & \\
        R_1 & 0 & 0 & \dots & -R_N \\
        1 & 1 & 1 & \dots & 1
    \end{bmatrix},
    &B_2 = 
    \begin{bmatrix}
        0 \\
        0 \\
        \vdots \\
        -1
    \end{bmatrix}.
\end{align}
Here, the last rows of $\Delta f(z)$, $A_{22}$, and $B_2$ enforce the current constraint, i.e.\ total current \(I\) is the sum of the elements in the branch current vector \(i\). The other entries in $\Delta f(z)$ and $A_{22}$ force equal terminal voltage for each cell, such that the sum of the OCV and ohmic voltage drop of any cell subtracted from another cell is equal to zero. With the DAE fully populated, (\ref{containsCurrents}) can be simplified to solve for the current through branch $x$,
\begin{equation}
i_x = \frac{R_\text{P}}{R_x}\left[\left(\sum_{k = 1}^{N} \frac{f(z_x) - f(z_k)}{R_k}\right) + I\right],
\label{branchCurrentOCV}
\end{equation}
where \(R_\text{P}\) is the equivalent parallel resistance of the whole module (the inverse of the sum of the admittances), and \(R_x\) and \(R_k\) are the individual total series resistances for branch \(x\) and branch \(k\) respectively. The index \(k\) refers to each individual branch from \(1\) to \(N\) contributing a term based on each branch voltage difference. This allows the model to be run without solving a DAE or inverting a matrix.. 

Equation (\ref{branchCurrentOCV}) may also be applied to the more complex RC-model shown in Fig.\ \ref{modelDiagram}. In this representation, each cell has three states: the voltage across the RC-pair, the SOC, and the core temperature rise. For our system with four cells in parallel, there are a total of twelve states. To account for the RC-pair voltage, we must subtract this from the respective OCV term. This results in a different numerator of equation (\ref{branchCurrentOCV}),
\begin{equation}
     i_x = \frac{R_\text{p}}{R_x}\left[\left(\sum_{k = 1}^{N} \frac{\left(f(z_x)-V_{\text{RC}_x}\right) - \left(f(z_k) - V_{\text{RC}_k}\right)}{R_k}\right) + I\right].
     \label{branchCurrent}
 \end{equation}
Defining total series resistance as $R_x = R_{{\Omega}_x}+R_{{\text{c}}_x}$ and the resistance in parallel with the capacitor of the RC-pair as $R'_x = R_{{\text{w}}_x} + R_{\text{ct}_x}$, for the $x$th parallel branch, \(V_{\text{RC}}\) obeys the following ODE: 
\begin{equation}
\label{dvdt_state_simplified}
     \dot{V}_{\text{RC}} = \frac{-V_{\text{RC}}}{R'C} + \frac{i}{C}.
 \end{equation}
Substituting (\ref{branchCurrent}) into (\ref{dvdt_state_simplified}), and continuing to omit the $x$ subscripts for clarity, for the $x$th branch we obtain: 
\begin{align}
\label{dvdt_state}
    \dot{V}_{\text{RC}} = &\frac{-V_{\text{RC}}}{R'C} + \\ \notag
    &\frac{R_\text{p}}{R C}\left[\left(\sum_{k = 1}^{N} \frac{\left(f(z)-V_{\text{RC}}\right) - \left(f(z_k) - V_{\text{RC}_k}\right)}{R_k}\right) + I\right],
\end{align}
while the SOC dynamics are described by
\begin{equation}
    \dot{z} = \frac{R_\text{p}}{R Q}\left[\left(\sum_{k = 1}^{N} \frac{\left(f(z)-V_{\text{RC}}\right) - \left(f(z_k) - V_{\text{RC}_k}\right)}{R_k}\right) + I\right]
    \label{dzdt_state}.
\end{equation}
In practice, $f(z)$ is implemented with a look-up table and linear interpolation over 520 OCV measurements. 

The equation for the thermal dynamics of the $x$th branch is the lumped thermal circuit in Fig.\ \ref{modelDiagram}: 
\begin{equation}
    \dot{T_\text{r}} = \frac{-T_\text{r}}{C_p(\text{Rth}_{\text{c-s}}+\text{Rth}_{\text{s-a}})} + \frac{Q_{\text{gen}}}{C_p}
\end{equation}
where \(C_p\) is the cell heat capacity, \(\text{Rth}_{\text{c-s}}\) is the thermal resistance from the core to the surface, and \(\text{Rth}_{\text{s-a}}\) is the thermal resistance from the surface to the ambient environment. The state $T$ is the temperature difference of the core with respect to ambient, i.e., $T  = T_\text{c} - T_\text{a}$. Heat generation in the cell is calculated with \(Q_{\text{gen}} = i^2R_\ohm + V_{\text{RC}}^2/(R_{\text{ct}}+R_\text{w})\), where it is assumed that the series resistor \(R_\ohm\) and the parallel total resistance $R' = R_{{\text{w}}} + R_{\text{ct}}$ generate heat in the cell. We ignore entropic heating here. The full thermal balance is then 
\begin{align}
\label{dTdt_state}
&\dot{T} = \frac{-T}{C_p(\text{Rth}_{\text{c-s}}+\text{Rth}_{\text{s-a}})} + \frac{V_{\text{RC}}^2}{C_p R'} +\\ \notag
&\frac{R_\ohm}{C_p} \cdot\left(\frac{R_{\text{p}}}{R_x}\sum_{k = 1}^{N} \frac{\left(f(z)-V_{\text{RC}_x}\right) - \left(f(z_k) - V_{\text{RC}_k}\right)}{R_k} + I\right)^2 .
\end{align}

Finally, the charge transfer resistance within the RC pair depends on the core temperature. This follows the Arrhenius relationship
\begin{equation}
R_{\text{ct}} = R_{\text{ct}_0}\exp\left(\frac{E_\text{a}}{\mathcal{R}}\left(\frac{1}{T_\text{c}} - \frac{1}{T_\text{a}}\right)\right),
\label{ArrheniusR}
\end{equation}
where \(R_{\text{ct}_0}\) is a reference resistance measured at ambient temperature \(T_\text{a}\), \(E_\text{a}\) is the activation energy for the reaction and $\mathcal{R}$ is the gas constant. Substituting this equation into the model couples the state equations together, because the branch currents depend on the internal temperature of each cell. The input \(u(t)\) to the model is the total current \(I(t)\), and the output \(y(t)\) is a vector of cell surface temperatures (only the $x$th row is given here), 
\begin{equation}
    u(t) = I(t) \ \text{ , } \ y(t) = T_\text{a} + T_\text{r}\left(\frac{\text{Rth}_{\text{s-a}}}{\text{Rth}_{\text{c-a}} + \text{Rth}_{\text{s-a}}}\right). 
\end{equation}

The model can be fitted to experimental current and temperature data and then used to gain insights into the maximum currents and temperatures in a parallel-connected module.

\subsection{Model Implementation}

The model was implemented with the \texttt{ode45} solver in Matlab, with initial SOC set at \(z=99.8 \%\), \(V_\text{RC}=0\), \(T = 0\), and \(T_\text{a}= 22.2 \,\celsius\), for all parallel cells respectively. At each time step, for each branch, $f(z)$ is calculated, the charge transfer resistance is updated according to the present temperature, and the vector of branch currents is calculated using 
\begin{equation}
    i = R_\text{p} Y \left(\left(V-V^T\right) Y+I\right),
\end{equation}
where scalar $R_\text{p}$ is as in (\ref{branchCurrent}), $Y = \begin{bmatrix} R_1^{-1} & R_2^{-1} & ... & R_N^{-1} \end{bmatrix}$ is a vector of admittances, $I$ is a scalar, and $V$ is a matrix of total voltage states $f(z)$ minus $V_{\text{RC}}$, 
\begin{equation}
    \begin{bmatrix}
        V_1 & \dots & V_1 \\
        V_2 & \dots & V_2 \\
        \vdots & & \vdots \\
        V_N & \dots & V_N
    \end{bmatrix}.
\end{equation}
Having established the branch currents, the model solves for the branch voltages, temperatures and SOCs at each time step using equations (\ref{dvdt_state}), (\ref{dzdt_state}), and (\ref{dTdt_state}).

\section{Experiments}
\label{sec:Experiments}

To probe the dynamics of imbalances between cells, an experiment (Fig.\ \ref{fig:moduleWiringDiagram}) was designed that investigated two fundamental cases: failure of a single cell and a failure of multiple cells. The purpose of the experiment is to establish a benchmark for the magnitude of the current and temperature imbalances and to show the validity of the model, fitting the model behaviour to the recorded dynamics. Data recorded in the single failure test configuration was first published in \cite{Lambert2026}. 

\subsection{Methodology}
\label{subsec:Methodology}

The first test case is the failure of a single cell in the module, characterised by an increased resistance between the failed cell and the DC power terminals. This case was intended to represent the resistance failures described in various articles \cite{BIRKL2017373, o2022lithium, KONG2018358, 9913923, Tanim2021} and discussed in Section \ref{sec:Intro}. Experimentally, this was achieved by adding a resistive busbar into the current path for a single cell in the module. The second failure case is an interconnection failure, where a resistor was added to the interconnect between each cell. This test case was used to investigate the effect of resistance imbalances such as corrosion or divergent degradation pathways \cite{ZWICKER2020100017, FailureReport, HENDRICKS2015113}. In addition to the two test cases, a baseline discharge test was also performed to provide a reference for the behaviour of the module with no added resistors. Due to the internal cell resistance-temperature dynamics, and safety considerations, it was not possible to emulate a true internal failure, however, the external arrangement was sufficient for model parameterisation and validation. Then further scenarios such as internal resistance or capacity imbalances were simulated using the model.
\begin{figure}%
    \includegraphics[width=0.5\textwidth, center]{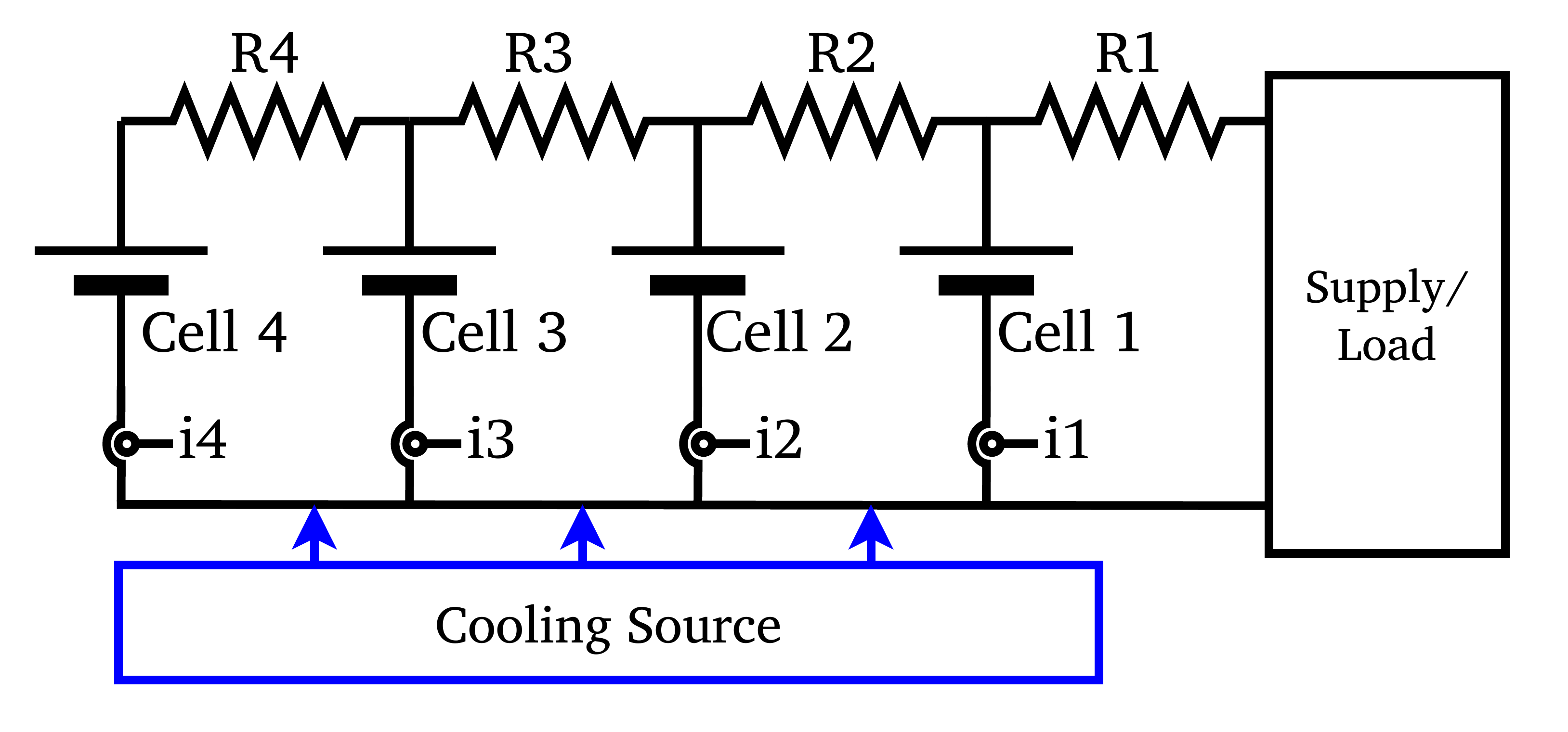}
    \caption{Module configuration with four parallel cells}
    \label{fig:moduleWiringDiagram}
\end{figure}

The test module was constructed with four 280 Ah LFP cells manufactured by RJ Energy. These were connected in parallel in the ladder configuration shown in Fig.\ \ref{fig:moduleWiringDiagram}, amounting to \SI{1120}{Ah} of energy storage. A \SI{20}{\micro\ohm} shunt current sensor was connected between the negative terminal of every cell and a single busbar that was used as the negative terminal of the module. On the positive side, the DC bus consisted of either a busbar or a resistive component connected between each cell to achieve different permutations of the test cases. The cells were placed in an enclosure with fans mounted on one side for cooling, replicating typical thermal management conditions for a grid storage module. A diagram of the test module is shown in Fig.\ \ref{fig:moduleWiringDiagram}. 

To achieve contact resistances similar to those of welded connections, busbars were cleaned with isopropyl alcohol, and an electrically conductive paste was applied to each connection surface. Parasitic resistances from busbar connections and other passive components were recorded using pulse measurements. The designed resistance configurations and actual measured values are shown in Table \ref{tab:R_Vals}. Resistances were selected based on a review of grid storage engineering practices and joining technologies \cite{SCHIMPE2018211, ZWICKER2020100017, RENIERS2023120774, 7854664}, with the intention that the test module be as similar as possible to a true grid storage battery pack. 

For each test, the procedure was as follows: (a) set the resistance values according to the required configuration in Table \ref{tab:R_Vals}; (b) charge module at 0.25C; (c) rest for 1 hour to reach equilibrium; (d) discharge at 0.45C (504 A) with a constant current until module terminal voltage reaches \SI{2.5}{V} or a cell safety limit is reached. During this process, individual cell voltages, currents, tab temperatures and body temperatures were measured. All cycling was performed using two ITECH IT6015C-80-450 bidirectional power supplies operating in parallel. 
\begin{table*}[h]
    \centering
    \begin{tabularx}{\textwidth}{|c|c||*{4}{>{\centering\arraybackslash}X}|*{4}{>{\centering\arraybackslash}X}|}
    \hline
    \multicolumn{2}{|c||}{\multirow{2}{*}{Test Configuration}} 
    & \multicolumn{4}{c|}{Designed Resistance Values (\SI{}{\micro\ohm})} 
    & \multicolumn{4}{c|}{Actual Resistance Values (\SI{}{\micro\ohm})}\\
    \cline{3-10}
    \multicolumn{2}{|c||}{} & R1 & R2 & R3 & R4 & R1 & R2 & R3 & R4 \\
    \hline
    \multicolumn{2}{|c||}{Baseline} & 0 & 0 & 0 & 0 & 15.8 & 12.0 & 16.6 & 17.1 \\
    \hline
    \multicolumn{2}{|c||}{Single Failure} & 0 & 0 & 0 & 250 & 15.5 & 12.5 & 16.3 & 264.2 \\
    \hline
    \multicolumn{2}{|c||}{Interconnect Failure} & 0 & 100 & 100 & 100 & 15.5 & 112.8 & 113.7 & 119.7 \\
    \hline
    \end{tabularx}
    \caption{Resistance values under the different test configurations.}
    \label{tab:R_Vals}
\end{table*}
Cell currents and voltages were recorded using Isabellenh{\"u}te IVT-S shunt sensors. The sensors were configured to send current and voltage samples at \SI{5}{Hz} to a Raspberry Pi datalogger. For safety reasons, this was also configured to control the power supply and interrupt the experiment if any measured data was beyond the manufacturer specified safety limits. This happened in both failure cases before the terminal voltage limits were reached, where the cell current exceeded the 280 A manufacturer limit \cite{RJenergyDatasheet} and an over-current fault stopped the tests. 

Temperature logging was implemented with a Picolog TC-08 datalogger with 8 channels of k-type thermocouples that measured both tab temperature and body temperature for all cells, respectively, at \SI{1}{Hz}. In the first 680 seconds of the interconnection failure test, the thermocouple on the tab of cell 1 showed periods of intermittent connection; these samples were removed from the data and replaced by interpolation over the disconnect periods. 

\subsection{Results}

\begin{figure*}[h!]
    \centering
    \begin{subfigure}[b]{0.32\textwidth}
        \begin{overpic}[width=\linewidth]{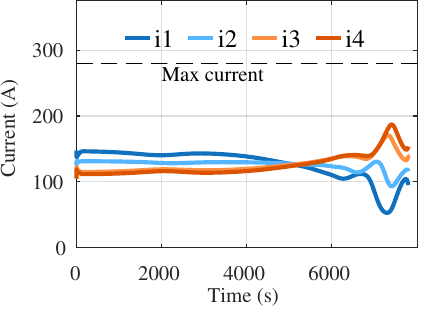}
        \put(3,72){\textbf{(a)}}
        \end{overpic}
        \label{fig:BTcurrExp}
    \end{subfigure}
    \hfill
    \begin{subfigure}[b]{0.32\textwidth}
        \begin{overpic}[width=\linewidth]{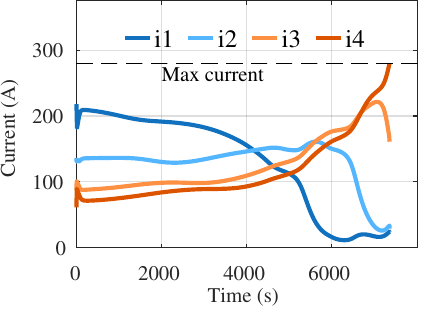}
        \put(3,72){\textbf{(b)}}
        \end{overpic}
        \label{fig:IFcurrExp}
    \end{subfigure}
    \hfill
    \begin{subfigure}[b]{0.32\textwidth}
        \begin{overpic}[width=\linewidth]{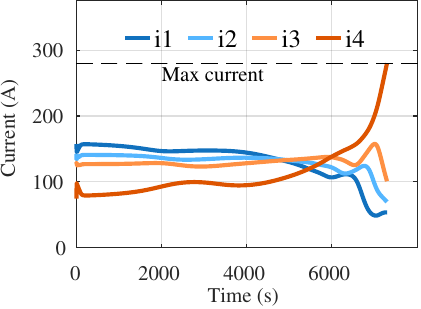}
        \put(3,72){\textbf{(c)}}
        \end{overpic}
        \label{fig:SFcurrExp}
    \end{subfigure}
    
    \vspace{-0.2cm} %
    
    \begin{subfigure}[b]{0.32\textwidth}
        \begin{overpic}[width=\linewidth]{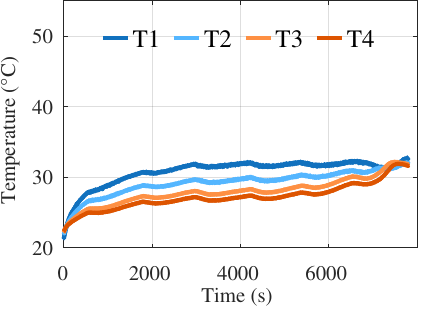}
        \put(3,72){\textbf{(d)}}
        \end{overpic}
        \label{fig:BTtempExp}
    \end{subfigure}
    \hfill
    \begin{subfigure}[b]{0.32\textwidth}
        \begin{overpic}[width=\linewidth]{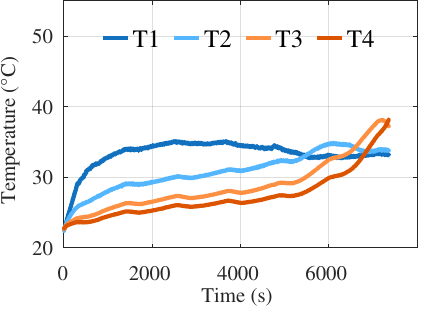}
        \put(3,72){\textbf{(e)}}
        \end{overpic}
        \label{fig:IFtempExp}
    \end{subfigure}
    \hfill
    \begin{subfigure}[b]{0.32\textwidth}
        \begin{overpic}[width=\linewidth]{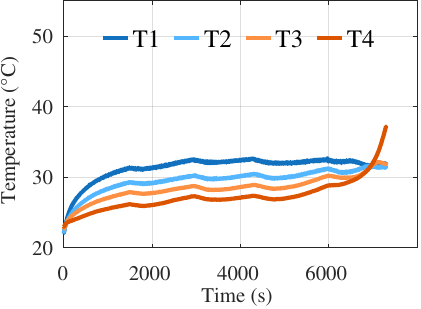}
        \put(3,72){\textbf{(f)}}
        \end{overpic}
        \label{fig:SFtempExp}
    \end{subfigure}
    \captionsetup{skip=-6pt}
    \caption{Current and temperature results from constant current discharge experiments. a) Baseline test currents. b) Interconnect failure currents. c) Single failure currents. d) Baseline test temperatures. e) Interconnect failure temperatures. f) Single failure temperatures.}
    \label{fig:expResults}
    \vspace{6mm}
\end{figure*}

The experimental measurements, shown in Fig.\ \ref{fig:expResults}, confirm a familiar literature result that modules with new cells, adequate thermal management, and welded (or equivalent) interconnects exhibit thermal gradients less than \SI{5}{\celsius}, as seen in the temperature data from the baseline test setup. The measurements also show that significant current and temperature differences can occur during cycling when there is a failure condition, with a maximum current difference of \SI{255}{A} and a maximum temperature difference of \SI{9}{\celsius} in the cell tabs during the cycle. A significant finding is the large scale of the currents at low SOC, where both failure tests had to be stopped due to cell current reaching the manufacturer recommended limit of \SI{280}{A}. The flat LFP OCV-SOC curve means that even for a large SOC difference between cells during cycling, the cell voltage is not sufficiently different to correct the SOC imbalance in the mid-SOC range. The module therefore experiences a rapid period of self-balancing as it discharges in the low SOC range, leading to extreme currents and temperatures. This phenomenon has also been suggested by other researchers \cite{weng2024current}. Not captured in the results is a potential extreme increase in temperature in the failed cell(s) at the end of the discharge. If the discharge had been left to continue, this temperature increase could have posed a safety risk, as evident from the rate of change in temperature in the `failed' cells. In theory, these extremes are not  detected by a conventional BMS if individual cell currents and temperatures are not recorded, which is often the case \cite{8626763}. 
  
\begin{figure*}[h!]
    \centering
    \begin{subfigure}[b]{0.32\textwidth}
        \begin{overpic}[width=\linewidth]{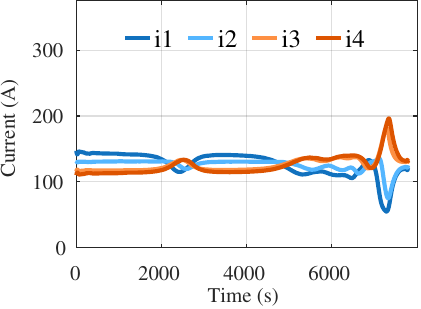}
        \put(3,72){\textbf{(a)}}
        \end{overpic}
        \label{fig:BTi}
    \end{subfigure}
    \hfill
    \begin{subfigure}[b]{0.32\textwidth}
        \begin{overpic}[width=\linewidth]{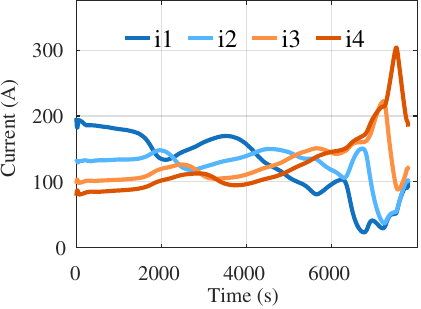}
        \put(3,72){\textbf{(b)}}
        \end{overpic}
        \label{fig:IFi}
    \end{subfigure}
    \hfill
    \begin{subfigure}[b]{0.32\textwidth}
        \begin{overpic}[width=\linewidth]{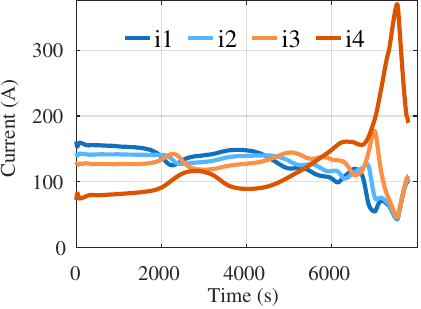}
        \put(3,72){\textbf{(c)}}
        \end{overpic}
        \label{fig:SFi}
    \end{subfigure}
    
    \vspace{-0.2cm} %
    
    \begin{subfigure}[b]{0.32\textwidth}
        \begin{overpic}[width=\linewidth]{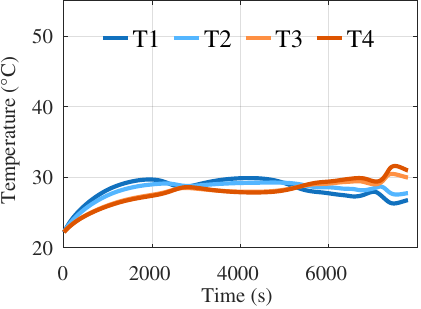}
        \put(3,72){\textbf{(d)}}
        \end{overpic}
        \label{fig:BTT}
    \end{subfigure}
    \hfill
    \begin{subfigure}[b]{0.32\textwidth}
        \begin{overpic}[width=\linewidth]{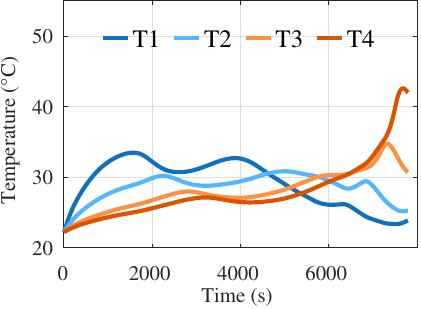}
        \put(3,72){\textbf{(e)}}
        \end{overpic}
        \label{fig:IFT}
    \end{subfigure}
    \hfill
    \begin{subfigure}[b]{0.32\textwidth}
        \begin{overpic}[width=\linewidth]{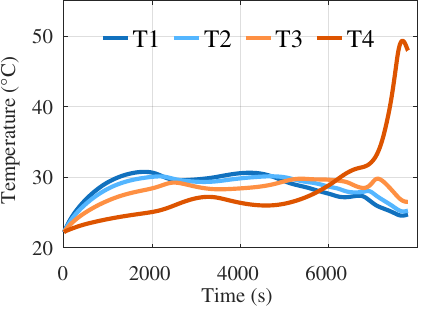}
        \put(3,72){\textbf{(f)}}
        \end{overpic}
        \label{fig:SFT}
    \end{subfigure}
    \captionsetup{skip=-6pt}
    \caption{Simulated cell currents and temperatures for model validation. a) Model baseline currents. b) Model interconnect currents. c) Model single failure currents. d) Model baseline temperatures. e) Model interconnect failure temperatures. f) Model single failure temperatures.}
    \label{fig:modelValidation}
    \vspace{4mm}
\end{figure*}

\begin{table*}[h!]
    \centering
    \begin{tabularx}{\textwidth}{|c||*{4}{>{\centering\arraybackslash}X}||*{4}{>{\centering\arraybackslash}X}|}
    \hline
    \multirow{2}{*}{Test Configuration} 
    & \multicolumn{4}{c||}{Cell Current RMSE (A)} 
    & \multicolumn{4}{c|}{Tab Temperature RMSE (\celsius)} \\
    \cline{2-9}
    & i1 & i2 & i3 & i4 & T1 & T2 & T3 & T4 \\
    \hline
    Baseline & 11.5 & 4.33 & 6.01 & 8.10 & 2.96 & 1.51 & 0.67 & 0.99 \\
    \hline
    Single Failure & 8.86 & 4.08 & 6.87 & 7.77 & 2.96 & 1.56 & 0.89 & 0.95 \\
    \hline
    Interconnect Failure & 34.8 & 22.6 & 19.0 & 17.8 & 4.68 & 2.86 & 1.74 & 0.69 \\
    \hline
    \end{tabularx}
    \caption{RMSE of the model vs. the experimental data in all test situations.}
    \label{tab:rmse_vals}
\end{table*}

\section{Model Parameterisation and Validation}

\subsection{Parameter estimation}

To ensure the model accurately represents the behaviour of a grid storage module, model parameters were estimated from measured cycling data. One challenge is that the thermal and electrical circuit parameters are coupled---the temperature-dependent equivalent circuit parameters require the thermal circuit to be accurate before the electrical parameters can be fitted correctly, and vice versa. Although all parameters could be fitted in a single optimisation problem, to reduce over-fitting the thermal model was decoupled and fitted separately to a single-state linear model, and then re-fitted once the electrical parameters were optimised. 

A simplified thermal model with one ODE was used to describe the increase in temperature in each cell, using measured current to quantify \(i^2R\) heat generation and the recorded temperature data shown in Fig.\ \ref{fig:expResults}. Here, \(i_x\) is the individual branch current recorded during testing, and \(R\) is the real part of the \SI{1}{Hz} cell impedance measured with individual electrochemical impedance spectroscopy (EIS) \cite{Barai2018}. To simplify the model, the same thermal parameters (heat capacity, thermal resistances) were used for all cells, which is a reasonable assumption for an industrial system with uniform thermal management \cite{SCHIMPE2018211}. Thermal parameters were fitted to tab temperature data, and for subsequent optimisation routines it was assumed that $\text{Rth}_{\text{c-s}}$ represented an in-plane thermal resistance from core to tab. Using this model, initial values for two thermal resistances plus a heat capacity were identified using the Matlab \texttt{ssest} function. 

Once the initial thermal parameters were estimated, the vector of electrical parameters $\hat{\theta} = \begin{bmatrix} \hat{R_\ohm} & \hat{R_\text{c}} & \hat{R_{\text{ct}_0}} & C & R_\text{w} & E_\text{a} \end{bmatrix}$ was estimated by minimising the following cost function using the Matlab \texttt{fmincon} nonlinear optimisation solver:  
\begin{align}
    & J(\hat{\theta}) = \\ \notag
    & \sqrt{(i_1(\hat{\theta}) - i_1^*)^2 + (i_2(\hat{\theta}) - i_2^*)^2 + (i_3(\hat{\theta}) - i_3^*)^2 + (i_4(\hat{\theta}) - i_4^*)^2}.
\end{align}
Here, \(i^*\) is the measured current and \(i(\hat{\theta})\) is the predicted current from the model with parameters \(\hat{\theta}\). Electrical parameters $\hat{R_\ohm}$, $\hat{R_\text{c}}$, and $\hat{R_{\text{ct}_0}}$ are vectors of four individual parameters (one for each cell), whereas \(R_\text{w}\), \(C\), and \(E_\text{a}\) are single values since the problem is not sensitive to these and the values are similar across all cells. Initial parameter values for the optimisation were set according to individual cell EIS measurements and the module resistance measurements discussed in section \ref{subsec:Methodology}. The initial activation energy \(E_\text{a}\) was set to \SI{67}{\kilo \joule \per \mol}, a value reported in the literature for the graphite/electrolyte interface \cite{Jow_2018}. Table \ref{tab:elecParams} lists the final optimised values of the parameters.

\begin{table}[htbp]
    \centering
    \begin{tabular}{|c|c|c|c|c|}
    \hline
        Param & Cell 1 & Cell 2 & Cell 3 & Cell 4 \\
        \hline
        $R_{\ohm}$ (\SI{}{\micro\ohm}) & 168.9 & 183.9 & 159.6 & 171.2 \\
        \hline
        $R_{\text{c}}$ (\SI{}{\micro\ohm}) & 127.8 & 150.9 & 218.2 & 225.9  \\
        \hline
        Q (\SI{}{\ampere\hour}) & 274.9 & 273.0 & 273.8 & 272.1 \\
        \hline
        $R_{\text{ct}_0}$ (\SI{}{\micro\ohm}) & 44.4 & 45.1 & 73.4 & 69.5 \\
        \hline
        $R_\text{w}$ (\SI{}{\micro\ohm}) & \multicolumn{4}{c|}{101} \\
        \hline
        $C$ (\SI{}{\mega\farad}) & \multicolumn{4}{c|}{4.5} \\
        \hline
        $E_\text{a}$ (\SI{}{\kilo \joule \per \mol}) & \multicolumn{4}{c|}{65} \\
        \hline
        $C_p$ (\SI{}{\joule \per \kelvin}) & \multicolumn{4}{c|}{205} \\
        \hline
        $\text{Rth}_{\text{c-s}}$ (\SI{}{\kelvin \per \watt}) & \multicolumn{4}{c|}{0.595} \\
        \hline
        $\text{Rth}_{\text{s-a}}$ (\SI{}{\kelvin \per \watt}) & \multicolumn{4}{c|}{1.362} \\
        \hline
    \end{tabular}
    \caption{Estimated thermal and electrical parameters}
    \label{tab:elecParams}
\end{table}

After electrical parameter fitting, thermal parameters \(\hat{\theta_{\text{th}}} = \begin{bmatrix} \text{Rth}_{\text{c-s}} & \text{Rth}_{\text{s-a}} & C_p \end{bmatrix}\) were then revisited and optimised using the same routine, minimising cost function
\begin{equation}
    J(\hat{\theta_{\text{th}}}) = \sqrt{(T(\hat{\theta_{\text{th}}}) - T^*)^2},
\end{equation}
where \(T^*\) represents the measured tab temperature and $T(\hat{\theta_{\text{th}}})$ is the tab temperature from the model execution with parameter set $\theta_{\text{th}}$. Since it was assumed all cells have the same thermal parameters, the cost function minimised the model error for one cell at a time, and in an effort to reduce the influence of external factors such as non-cell heat sources, the parameters from the cell with the best fit to the input/output data were selected. Full results are shown in Table \ref{tab:elecParams}.

\subsection{Model validation}

Since there was a limited amount of experimental data, the model parameters were estimated from the baseline test data and validated against the failure cases. This approach reserved twice as much data for validation to ensure that the model was consistent under different cycling scenarios. Model validation data was generated by adding the connection resistance values in Table \ref{tab:R_Vals} to the \(R_c\) parameters while keeping all other parameters constant. The same constant-current discharge tests were simulated at 0.45C, and the results displayed in Fig.\ \ref{fig:modelValidation} show how the model performed compared to the experimental data shown in Section \ref{sec:Experiments}. 

The fitted model is able to replicate the dynamics of the system, and is particularly useful in the areas of rapid temperature increase. On the interconnect failure, at the time when the experiment was stopped, the difference in tab temperature between the model prediction and the experimental measurement was \SI{2.4}{\celsius} and \SI{1.5}{\celsius}, for cells three and four respectively. For the single failure, the difference in temperature in the failed cell between the model and the experiment was \SI{0.3} {\celsius} when the experiment was halted. These results show that at the critical points during the cycle when cell temperature is increasing, the temperature model was accurate to within 7\%. Full RMSE values for the currents and temperatures of all tests are recorded in table \ref{tab:rmse_vals}.

The error in the model is due to both the experimental setup and the unmodeled internal behaviour of the cells. In Fig.\ \ref{fig:modelValidation}, the model currents have greater fluctuations in the middle of the cycle, and also in the low SOC range. These fluctuations are related to the shape of the OCV curve, which has features or `bumps' at these SOC points, most likely due to graphite staging \cite{doi:10.1021/acs.chemmater.2c01976}. When calculating currents with equation (\ref{branchCurrent}), the voltage delta ($f(z)$) terms in the model are higher around these SOC regions, and therefore dominate the equation. However, these features in the OCV curve are not as significant when the cells are under sustained load. A possible explanation is the charge heterogeneity internal to large format cells that results from multi-scale nonuniformity on the material level \cite{C9TA06977A}. Charge heterogeneity has the overall effect of smoothing the OCV curve \cite{Lin2022}, resulting in smaller voltage differences between parallel-connected cells. Furthermore, the accuracy of the thermal model compared to the experiment is affected by heat generation in the passive components in the circuit, which is also not captured in the model. During the experiment, two \SI{120}{\milli\meter\squared} cables were connected in parallel at each terminal to connect the module DC bus to the power supply. These cables were heated during cycling, and injected heat into the system at the busbar connection of cell one. For this reason, cell one heated up more than the model prediction, and did not cool down when the current dropped. It is assumed that in DC power systems where the current demand is high, the conductors in the power path do not transfer additional heat to the cells. This issue was mitigated by fitting the model to the temperature dynamics of cell three, as it was located sufficiently far from the cable heating. 

Using the model to extrapolate the results of the failure cases, internal thermal gradients reach maxima of \SI{29}{\celsius} and \SI{38}{\celsius} peak-to-peak in the interconnect failure and the single failure case, respectively. Furthermore, maximum internal cell temperatures were \SI{51}{\celsius} and \SI{61}{\celsius}, according to the model in these respective examples. 

\section{System Analysis}

\begin{figure}[t]
    \centering
    \includegraphics[width=0.45\textwidth]{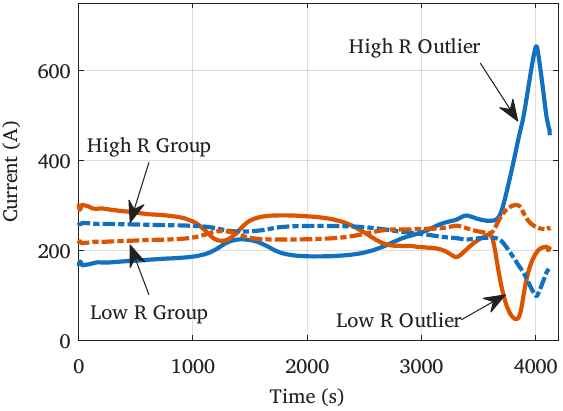}
    \caption{Currents for two simulations: module with a high resistance outlier (blue) and a module with a low resistance outlier (orange).}
    \label{fig:R_outliers_i}
\end{figure}

\begin{figure}[h]
    \centering
    \includegraphics[width=0.45\textwidth]{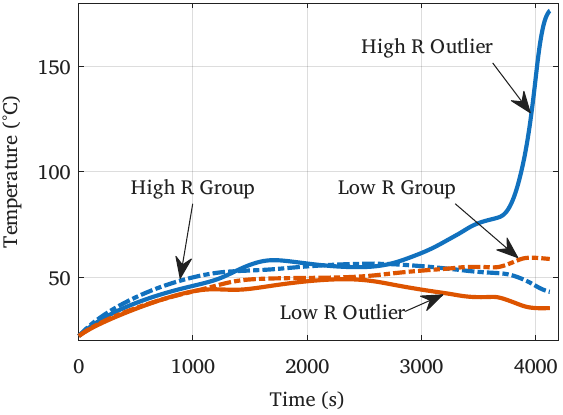}
    \caption{Temperatures for two sims: module with a high resistance outlier (blue) and a module with a low resistance outlier (orange).}
    \label{fig:R_outliers_T}
\end{figure}

\begin{figure}[t]
    \centering
    \includegraphics[width=0.45\textwidth]{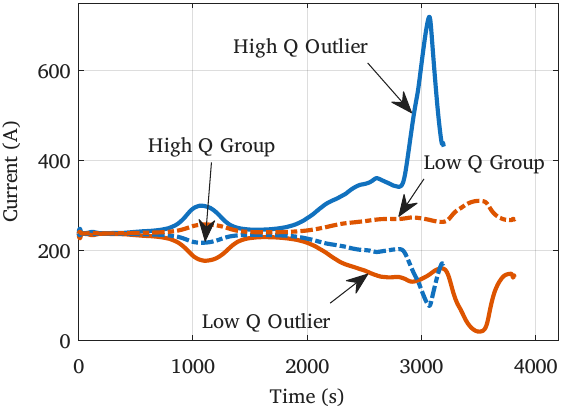}
    \caption{Currents for two simulations: module with a high capacity outlier (blue) and a module with a low capacity outlier (orange).}
    \label{fig:Q_outliers_i}
\end{figure}

\begin{figure}[h]
    \centering
    \includegraphics[width=0.45\textwidth]{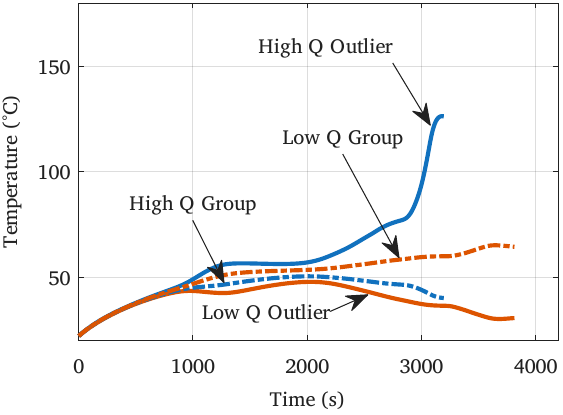}
    \caption{Temperatures for two sims: module with a high capacity outlier (blue) and a module with a low capacity outlier (orange).}
    \label{fig:Q_outliers_T}
\end{figure}

Using the model, various simulation studies were performed to investigate how differences in internal resistance and capacity impact the current and temperature in parallel modules. In the first study, four chosen parameter sets with different outliers were simulated: (1) a high resistance case where the outlier has double the average value of $R_{\ohm}$ from Table \ref{tab:elecParams}, (2) a low resistance outlier case where one cell has half the average value, (3) a low capacity outlier case where one cell has 70\% the nominal capacity, and (4) a high capacity outlier case where three cells are at 70\% of nominal capacity. The remaining cells were assumed to be identical and have the averaged parameters of Table \ref{tab:elecParams}. The resulting cell currents and internal temperatures are shown in Figs. \ref{fig:R_outliers_i}--\ref{fig:Q_outliers_T}. The simulations discharged the module at 952 A to ensure that the imbalances were significant and to emulate a one-hour grid storage load profile (although simulations with lower module capacity took less than 1 hour). In Figs. \ref{fig:R_outliers_i}--\ref{fig:Q_outliers_T} the dashed lines show three identical current/temperature trajectories superimposed, labelled the `Group' dynamics of the cells excluding the outlier cell. The solid lines show a single current or temperature for the outlier cell with either the high parameter case in blue, or the low parameter case, in orange. 
    
In these examples, resistance differences result in an immediate difference in load current, according to the resistance ratio term in equation (\ref{branchCurrent}). This leads to a temporary SOC imbalance that is corrected with the rebalancing current at the end of the cycle. Conversely, a capacity imbalance does not immediately cause the cells to have different currents, but over time, equal loading in cells with different capacities creates an SOC imbalance. The SOC imbalance is corrected in the second half of the cycle, as the cell(s) at a higher SOC and voltage are loaded with a greater current, according to equation (\ref{branchCurrent}). The finding from these examples is that the module will be more prone to high currents and temperatures during discharge when the outlier cell discharges slower than the group. This maximises the number of voltage delta terms in equation (\ref{branchCurrent}) that are the same sign as at the total current, and therefore results in a large cell current. Assuming that as cells degrade, resistance  increases monotonically, and capacity decreases monotonically, a module with more cells in parallel will be more robust to low capacity degradation outliers, and less robust to high resistance degradation outliers. Therefore, careful consideration should be given to the number of cells in parallel on the basis of the expected heterogeneity within a BESS module.  

\subsection{Temperature sensitivity analysis}

While these findings are insightful, it is inefficient to continue a case-by-case approach on every possible parameter set. Instead we analysed the sensitivity of the module thermal differences with a global method, Sobol indices, that quantifies how much the total variance in temperature differences results from the variance of a single parameter. First-order and total-effect Sobol indices were calculated according to the methods in \cite{SALTELLI2010259}. This sensitivity analysis method is reliable option for nonlinear models \cite{YANG2011444}, and has been applied to battery models successfully in the past in \cite{DANGWAL20237120}. 

\begin{figure}
    \includegraphics[width=0.45\textwidth, center]{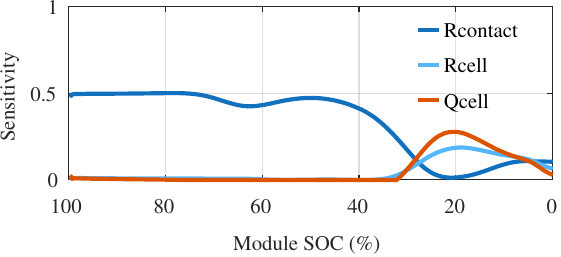}
    \caption{First order Sobol.}
    \label{fig:Sobol_FO}
\end{figure}

\begin{figure}%
    \includegraphics[width=0.45\textwidth, center]{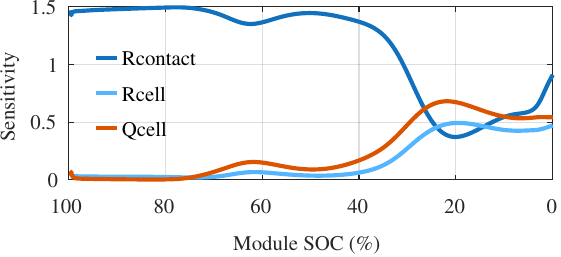}
    \caption{Total effect Sobol.}
    \label{fig:Sobol_TF}
\end{figure}

The indices were calculated on a twelve-dimensional parameter space, with the contact resistance, ohmic resistance, and capacity of all four cells were included in the analysis. The first-order indices and total-effect indices were calculated using simulated discharge output data from a scrambled Sobol sequence of 262,144 parameter sets, then combined into three indices, respectively for contact resistance, cell resistance, and capacity. The parameter space used in this analysis is shown in Table \ref{tab:sobolRange}, and results are shown in Figs. \ref{fig:Sobol_FO} and \ref{fig:Sobol_TF}. Note that in Fig.\ \ref{fig:Sobol_TF} the total index exceeds 1 because the twelve dimensional space was compressed into a three dimensional space by summing the indices associated with the same parameter on different cells.  

\begin{table}[]
    \centering
    \begin{tabular}{|c|c|c|}
    \hline
        $R_\text{c}$ (\SI{}{\micro\ohm}) & $R_\ohm$ (\SI{}{\micro\ohm})& $Q$ (\SI{}{\ampere\hour}) \\
        \hline
        124-424 & 172-344 & 191-273 \\
        \hline
    \end{tabular}
    \caption{Range of each parameter in the Sobol calculations}
    \label{tab:sobolRange}
\end{table}

Figs. \ref{fig:Sobol_FO} and \ref{fig:Sobol_TF} show how the parameter sensitivities develop during a discharge simulation. At first, the thermal difference depends mostly on contact resistance, and then towards the end of the discharge it is more dependent on cell capacity and resistance. The impact of contact resistance is in part due to the larger range of the contact resistances, because of the effect of interconnect resistances in the ladder configuration. Even if the interconnect resistance is small for each cell, the additive nature results in large thermal imbalances. For instance \SI{15}{\micro\ohm} of resistance at each interconnect can cause thermal gradients of \SI{25}{\celsius} and maximum temperatures of \SI{68}{\celsius}, when the module is discharged at 0.85C. From this analysis, it is assumed that variations in internal resistance and contact resistance pose the greatest risk to the parallel module, for the range of capacity and resistance variation defined in Table \ref{tab:sobolRange}.

\subsection{Derivation of safety limits}

To consolidate the findings of this study, safety thresholds for maximum parameter variability between cells have been derived using a constrained optimisation scheme. A cost function was set up to maximise the difference between a single parameter in the model and the average of that parameter in the remaining cells, subject to a temperature constraint. The temperature constraint assumed that the cell core temperature must be below \SI{60}{\celsius} to be considered safe \cite{FENG2018246, SPOTNITZ200381, RJenergyDatasheet}. To simulate degradation-related variability, the resistance was increased in its deviation from the mean and the capacity decreased. The thresholds are displayed in Table \ref{tab:safetyLims} for a 0.45C discharge and 0.85C discharge respectively. The limits are normalised to the mean parameter in the remaining cells, so that the maximum variability for each parameter \(\theta\) is expressed as: 

\begin{equation}
    \frac{|\theta_{\text{limit}} - \theta_{\text{mean}}|}{\theta_{\text{mean}}}\times100\%
\end{equation}

\begin{table}[]
    \centering
    \begin{tabular}{|c|c|c|}
    \hline
        Parameter & 0.45C Limit & 0.85C Limit \\
        \hline
        $R_\ohm$ & 97.3\% & 11.2\% \\
        \hline
        $R_\text{c}$ & 243.9 \% & 22.3\%\\
        \hline
        $R_{\text{ct}_0}$ & 1391\% & 432.3\% \\
        \hline
        $Q$ & N/A & 16.4\%\\
        \hline
    \end{tabular}
    \caption{Maximum allowable percentage change in each parameter to ensure maximum temperature of \SI{60}{\celsius} at two C-rates.}
    \label{tab:safetyLims}
\end{table}

A lower percentage signifies less room for parameter variability, and lower system robustness. Notable in these results is the tightening of the variability threshold at the higher C-rate. Additionally, in this configuration no amount of capacity heterogeneity is able to heat a cell above \SI{60}{\celsius} at 0.45C. 

\subsection{Thresholds for safe operation}
\label{subsec:robust}

\begin{figure}[h]
    \includegraphics[width=0.45\textwidth, center]{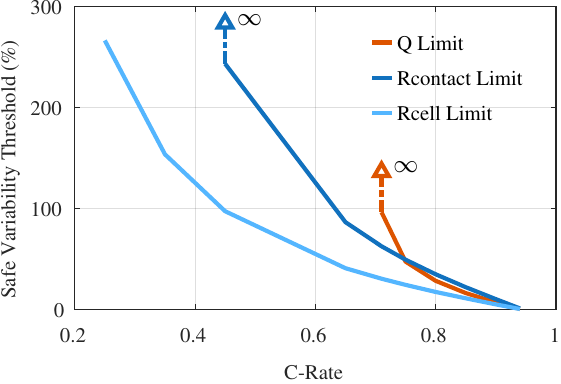}
    \caption{Safety thresholds vs. C-rate at 100\% DoD, 4 cells.}
    \label{fig:crateRobustness}
\end{figure}

\begin{figure}[h]
    \includegraphics[width=0.45\textwidth, center]{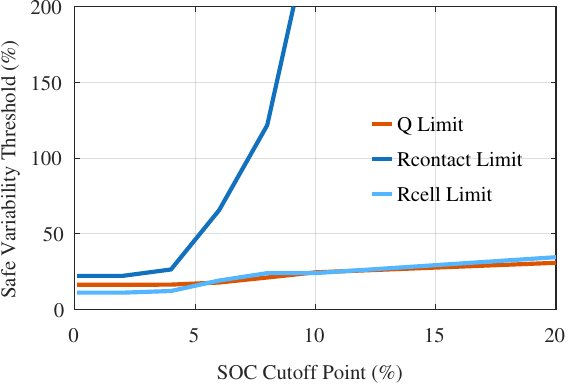}
    \caption{Safety thresholds vs.\ SOC cut-off point at 0.85C, 4 cells.}
    \label{fig:SocRobustness}
\end{figure}

\begin{figure}[h]
    \includegraphics[width=0.45\textwidth, center]{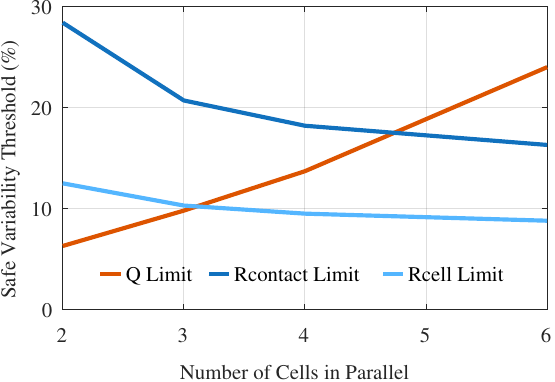}
    \caption{Safety thresholds vs.\ number of cells at 0.85C, 100\% DoD.}
    \label{fig:noCellsRobustness}
\end{figure}

In an effort to understand what actions can be taken to mitigate the risk of overheating in parallel modules, the safety thresholds were plotted for different C-rates, depths of discharge (DoD), and number of cells in parallel. This analysis shows how the safe set of parameter values expands based on different use cases. By understanding these factors, BESS integrators can maximise the performance of a system while also being confident in the safety of the system. The resulting thresholds are shown in Figs. \ref{fig:crateRobustness}--\ref{fig:noCellsRobustness}.

An intuitive result shown in Fig.\ \ref{fig:SocRobustness} is that avoiding cycling in regions where the OCV-curve is steep, such as in the low SOC range, avoids the large rebalancing currents discussed throughout the paper, and increases the defined safety thresholds. Additionally, in Fig.\ \ref{fig:crateRobustness}, the area below or to the left of each respective curve is the `safe' parameter space where there is no level of heterogeneity in the indicated parameters that can lead to temperatures above \SI{60}{\celsius}. Although operating in this guaranteed `safe-zone' is an attractive idea, it would require a major compromise on pack performance, where the module can only be cycled at a lower rate and in a restrictive SOC window. Instead the robustness metrics offer insights into strategies that maximise performance while decreasing risk to an acceptable level. For example, strategically reducing the current in the low SOC range, where the slope of the OCV function is steepest, can improve system safety without making a significant compromise on accessible energy. 

Although these results were generated for a specific module configuration, they can be generalised. The global sensitivity analysis is a relative metric based on simulations with random parameter sets sampled from the ranges in Table \ref{tab:sobolRange}. These simulations are not tied to the experimental module, but only to the parameter space defined from the literature. The robustness metrics are normalised, and the observed trends arise from the model dynamics in equations \eqref{dvdt_state}–\eqref{dTdt_state}. Therefore, the trends in Figs. \ref{fig:crateRobustness}–\ref{fig:noCellsRobustness} are broadly applicable across parallel cell groups.

\section{Conclusion}

This paper realistically measured and simulated the scale of possible temperature imbalances that could occur in large-format parallel-connected cells such as those used in grid storage systems. A computationally efficient modelling framework was presented that successfully replicates the measured electrical and thermal behaviour of parallel cells. Along with the modelling framework, this paper presents a method for analysing the system sensitivity to parameter variations between cells, and an optimisation scheme to derive limits on cell-to-cell variability. The overarching contribution is a technique for the design and control of a parallel pack to ensure safety considering the sensing limitations of industrial systems. 

Experiments have shown that in modules with new cells and adequate thermal management, it is possible to contain thermal imbalances to less than \SI{5}{\celsius} during routine cycling. However, interconnection resistance increases or cells with accelerated degradation can lead to thermal imbalances above \SI{25}{\celsius}. Furthermore, if individual cell temperatures are not monitored by the BMS, heterogeneity can cause failures associated with overheating due to extreme rebalancing currents. This could be a significant issue for second-life pack manufacturers who are combining cells with differing parameters into modules. 

To mitigate the risks of overheating, parallel modules can be made more robust to the impacts of variability if the maximum C-rate is reduced and also if the SOC window is reduced. Although both mitigation strategies compromise performance, this poses an emerging optimisation question, that is, what is the best way to get the most out of the pack while maintaining a high level of confidence in the safety of the system as it experiences degradation? This optimisation question is not straight forward, as there are several ways that cells in a module can differ from one another, and multiple different ways to increase the size of the safe parameter space. Understanding the mechanisms that impact current and temperature imbalance can facilitate designs that balance pack performance and safety.

\section*{Acknowledgements}
\noindent
This work has been supported by Brill Power and the Royal Commission for the Exhibition of 1851.

\section*{Data and code availability}
\noindent
The code used in this study will be made publicly available upon publication of the article. Data supporting findings in this study is made available upon request.

\section*{Competing Interests}
\noindent
JR, EC, and DF are employed by Brill Power. DH is a cofounder of Brill Power. SD has no competing interests. 

\bibliographystyle{elsarticle-num}

\end{document}